\documentclass[twocolumn,twocolappendix]{aastex631}

\usepackage{mathrsfs}

\newcommand{\pfrac}{$P_{\text{frac}}$}

\newcommand{\ttcooz}{\mbox{$^{13}$CO(1--0)}}
\newcommand{\Ngas}{$N_{\text{gas}}$}
\newcommand{\smallngas}{$n_{\text{gas}}$}
\newcommand{\NHt}{$N_{\text{H$_2$}}$}
\newcommand{\NHe}{$N_{\text{He}}$}
\newcommand{\muHt}{$\mu_{\text{H$_2$}}$}
\newcommand{\mup}{$\mu_p$}
\newcommand{\kms}     {km\,s$^{-1}$}

\shorttitle{G47}
\shortauthors{Stephens et al.}
\begin{document}

\title{The Magnetic Field in the Milky Way Filamentary Bone G47}

\author{Ian W. Stephens}
\affiliation{Department of Earth, Environment, and Physics, Worcester State University, Worcester, MA 01602, USA \href{mailto:istephens@worcester.edu}{istephens@worcester.edu}}
\affiliation{Center for Astrophysics $|$ Harvard \& Smithsonian, 60 Garden Street, Cambridge, MA 02138, USA} 

\author{Philip C. Myers}
\affiliation{Center for Astrophysics $|$ Harvard \& Smithsonian, 60 Garden Street, Cambridge, MA 02138, USA} 

\author{Catherine Zucker}
\altaffiliation{Hubble Fellow}
\affiliation{Space Telescope Science Institute, 3700 San Martin Drive, Baltimore, MD 21218, USA}
\affiliation{Center for Astrophysics $|$ Harvard \& Smithsonian, 60 Garden Street, Cambridge, MA 02138, USA} 

\author{James M. Jackson}
\affiliation{SOFIA Science Center, USRA, NASA Ames Research Center, Moffett Field CA 94045, USA}

\author{B-G Andersson}
\affiliation{SOFIA Science Center, USRA, NASA Ames Research Center, Moffett Field CA 94045, USA}

\author{Rowan Smith}
\affiliation{Jodrell Bank Centre for Astrophysics, Department of Physics and Astronomy, University of Manchester, Oxford Road, Manchester, M13 9PL, UK}

\author{Archana Soam}
\affiliation{SOFIA Science Center, USRA, NASA Ames Research Center, Moffett Field CA 94045, USA}
\affiliation{Indian Institute of Astrophysics, II Block, Koramangala, Bengaluru 560034, India}

\author{Cara Battersby}
\affiliation{University of Connecticut, Department of Physics, 196A Auditorium Road, Unit 3046, Storrs, CT 06269, USA}

\author{Patricio Sanhueza}
\affiliation{National Astronomical Observatory of Japan, National Institute of Natural Sciences, 2-21-1 Osawa, Mitaka, Tokyo 181-8588, Japan}
\affiliation{Department of Astronomical Science, SOKENDAI (The Graduate University for Advanced Studies), 2-21-1 Osawa, Mitaka, Tokyo 181-8588, Japan}

\author{Taylor Hogge}
\affiliation{Institute for Astrophysical Research, Boston University, 725 Commonwealth Avenue, Boston MA 02215, USA}

\author{Howard A. Smith}
\affiliation{Center for Astrophysics $|$ Harvard \& Smithsonian, 60 Garden Street, Cambridge, MA 02138, USA} 

\author{Giles Novak}
\affiliation{Center for Interdisciplinary Exploration and Research in Astrophysics (CIERA), and Department of Physics \& Astronomy, Northwestern University, 2145 Sheridan
Rd., Evanston, IL 60208, USA}

\author{Sarah Sadavoy}
\affiliation{Department of Physics, Engineering and Astronomy, Queen's University, 64 Bader Lane, Kingston, ON, K7L 3N6, Canada}

\author{Thushara Pillai}
\affiliation{Institute for Astrophysical Research, Boston University, 725 Commonwealth Avenue, Boston MA 02215, USA}

\author{Zhi-Yun Li}
\affiliation{Astronomy Department, University of Virginia, Charlottesville, VA 22904, USA}

\author{Leslie W. Looney}
\affiliation{Department of Astronomy, University of Illinois, 1002 West Green Street, Urbana, IL 61801, USA}

\author{Koji Sugitani}
\affiliation{Graduate School of Science, Nagoya City University, Mizuho-ku, Nagoya, Aichi 467-8501, Japan}

\author{Simon Coud\'e}
\affiliation{SOFIA Science Center, USRA, NASA Ames Research Center, Moffett Field CA 94045, USA}

\author{Andr\'es Guzm\'an}
\affiliation{National Astronomical Observatory of Japan, National Institute of Natural Sciences, 2-21-1 Osawa, Mitaka, Tokyo 181-8588, Japan}

\author{Alyssa Goodman}
\affiliation{Center for Astrophysics $|$ Harvard \& Smithsonian, 60 Garden Street, Cambridge, MA 02138, USA}

\author{Takayoshi Kusune}
\affiliation{Graduate School of Science, Nagoya University, Chikusa-ku, Nagoya 464-8602, Japan}

\author{F\'abio P. Santos}
\affiliation{Max-Planck-Institute for Astronomy, K\"onigstuhl 17, D-69117 Heidelberg, Germany}

\author{Leah Zuckerman}
\affiliation{Center for Astrophysics $|$ Harvard \& Smithsonian, 60 Garden Street, Cambridge, MA 02138, USA} 

\author{Frankie Encalada}
\affiliation{Department of Astronomy, University of Illinois, 1002 West Green Street, Urbana, IL 61801, USA}


\begin{abstract}
Star formation primarily occurs in filaments where magnetic fields are expected to be dynamically important. The largest and densest filaments trace spiral structure within galaxies. Over a dozen of these dense ($\sim$10$^4$\,cm$^{-3}$) and long ($>$10\,pc) filaments have been found within the Milky Way, and they are often referred to as ``bones." Until now, none of these bones have had their magnetic field resolved and mapped in their entirety. We introduce the SOFIA legacy project FIELDMAPS which has begun mapping $\sim$10 of these Milky Way bones using the HAWC+ instrument at 214\,$\mu$m and 18$\farcs$2 resolution. Here we present a first result from this survey on the $\sim$60\,pc long bone G47. Contrary to some studies of dense filaments in the Galactic plane, we find that the magnetic field is often not perpendicular to the spine (i.e., the center-line of the bone). Fields tend to be perpendicular in the densest areas of active star formation and more parallel or random in other areas. The average field is neither parallel or perpendicular to the Galactic plane nor the bone. The magnetic field strengths along the spine typically vary from $\sim$20 to $\sim$100\,$\mu$G. Magnetic fields tend to be strong enough to suppress collapse along much of the bone, but for areas that are most active in star formation, the fields are notably less able to resist gravitational collapse.
\end{abstract}


\section{Introduction} \label{sec:intro}
High-mass star-forming molecular clouds in spiral galaxies primarily follow the spiral arms. As such, these molecular clouds and their young stellar objects (YSOs) are used to trace spiral structure within the Milky Way \citep[e.g,.][]{Reid2014}. Observations from $Spitzer$ revealed that some of these star-forming clouds are dense ($\sim$10$^4$\,cm$^{-3}$), high-mass, and exceptionally elongated (e.g., over 80\,pc~$\times$~0.5\,pc for the Nessie filament; \citealt{Jackson2010,Goodman2014}). These filamentary structures are called bones because they delineate the densest parts of arms in a spiral galaxy, just as bones delineate the densest parts of arms in a human skeleton \citep{Goodman2014}. \citet{Zucker2015,Zucker2018b} identified 18 bone candidates in the Milky Way using strict criteria: they must be velocity coherent along the structure, have aspect ratios of $>$50:1, lie within 20\,pc of the Galactic plane, and lie mostly parallel to the Galactic plane.  The physical properties of these bones are well-characterized, including measurements of lengths, widths, aspect ratios, masses, column densities, dust temperatures, Galactic altitudes, kinematic separation from arms in $l-v$ space, and distances \citep{Zucker2015,Zucker2018b}.  However, the magnetic field (henceforth, B-field), which can potentially support the clouds against gravitational collapse or guide mass flow, has been mostly unconstrained for bones. 

Since non-spherical dust grains align with their short axis along the direction of the B-field, thermal dust emission is polarized perpendicular to the B-field \citep[e.g.,][]{Andersson2015}. Consequently, in star-forming clouds, polarimetric observations at (sub)millimeter wavelengths are the most common way to constrain the B-field morphology. \citet{Pillai2015} used the James Clerk Maxwell Telescope (JCMT) SCUBAPOL polarimetric observations at 20$\arcsec$ (0.3 pc) resolution to constrain the B-field morphology of a small, bright section of the bone G11.11--0.12 (also known as the Snake). They found that the field toward this section is perpendicular to the bone, and they estimated the B-field to be $\sim$300\,$\mu$G and found a mass-to-flux parameter that is approximately unstable to gravitational collapse. Until this work, these observations were the only published measurements of the field morphology of part of a bone at these scales. However, other studies have probed B-fields in shorter, high-mass filamentary structures, such as G35.39--0.33  \citep{Liu2018,Juvela2018}, NGC~6334 \citep{Arzoumanian2021}, and  G34.43+0.24 \citep{Soam2019}. In general, these studies found that the field is perpendicular to the filament (i.e., elongated dense clouds) in their densest regions and parallel in the less dense regions, such that the parallel fields may feed material into the denser regions of the filament.  Moreover, the B-fields may provide some support against collapse. One much smaller scales (1000 -- 10000\,au), YSOs themselves can have diverse magnetic field morphologies such as spiral-like, hourglass, and radial \citep[e.g., as seen in the MAGMAR survey;][]{Cortes2021,FL2021,Sanhueza2021}. Focusing on the large-scale observations of filaments, it is important to establish if fields are universally perpendicular to the spines of the main filament and whether the B-field strength is sufficient to help support the filament from collapse. As such, polarization maps of the largest filamentary structures, i.e., the bones, will be one of the best ways to investigate field alignment with filamentary structures. Such observations also constrain the importance of magnetic fields for star formation within spiral arms. Based on polarization observations of face-on spiral galaxies, the inferred large-scale field appears to be along spiral arms \citep{LiHenning2011,Beck2015}.

In a legacy project called FIlaments Extremely Long and Dark: a MAgnetic Polarization Survey (FIELDMAPS), we are using the High-resolution Airborne Wideband Camera Plus (HAWC+) polarimeter \citep{Dowell2010,Harper2018} on the Stratospheric Observatory for Infrared Astronomy (SOFIA) to map 214\,$\mu$m polarized dust emission across $\sim$10 of the 18 known bones. This survey is currently in progress, and this Letter focuses on the early results for the bone G47.06+0.26 (henceforth, G47). The HAWC+ polarimetric maps represent the most detailed probe of the B-field morphology across an entire bone to date. The resolution of $Planck$ is too coarse ($\sim$10$\arcmin$, e.g., \citealt{Planck35}) to resolve any bones.

The kinematic distance to G47 is 4.4\,kpc \citep{Wang2015}, but based on a Bayesian distance calculator from \citet{Reid2016} and its close proximity to the Sagittarius Far Arm, 
\citet{Zucker2018b} determined that the more likely distance is 6.6\,kpc, which we adopt. \citet{Zucker2018b} examined the physical properties of G47 in detail. 
The bone has a length of 59\,pc and a width of $\sim$1.6\,pc. The median dust temperature is $T_{\text{dust}}$\,=\,18\,K and the median H$_2$ column density is $N_{\text{H$_2$}}$\,=\,4.2\,$\times$10$^{21}$\,cm$^{-3}$. The total mass of the bone is 2.8\,$\times$10$^{4}$\,$M_\odot$, and the linear mass density is 483~$M_\odot$\,pc$^{-1}$. \citet{Xu2018} analyzed the kinematics of G47. Among their results, they found that the linear mass density is likely less than the critical mass density to be gravitationally bound, and suggested external pressure may help support the bone from dispersing under turbulence. They also found a velocity gradient across the width (but not the length) of G47, which may be due to the formation and growth of G47. In this Letter we analyze the inferred B-field morphology in G47 as mapped by SOFIA HAWC+.



\section{Observations and Ancillary Data} \label{sec:obs}
\subsection{SOFIA HAWC+ Observations}
G47 was observed with SOFIA HAWC+ in Band E, which is centered at 214\,$\mu$m and provides a resolution of 18$\farcs$2 \citep{Harper2018} or 0.58 pc resolution at a distance of 6.6 kpc. The observations were taken over multiple flights in September 2020 during the OC8E HAWC+ flight series as part of the FIELDMAPS legacy project. The polarimeter's field of view is 4$\farcm$2\,$\times$\,6$\farcm$2. The entire bone was mapped by mosaicking together four separate on-the-fly (SCANPOL) maps. 
The total time on source for the combined observations was 4070\,s. We use the Level 4 delivered products from the SOFIA archive.\footnote{\url{https://irsa.ipac.caltech.edu/applications/sofia/}} The pixel sizes are $3\farcs7 \times 3\farcs7$, which oversamples the 18$\farcs$2 beam. Errors along the bone for the Stokes~$I$, $Q$, and $U$ maps varied from about 0.5 to 0.8\,mJy\,pixel$^{-1}$. From the Stokes parameters, the polarization angle, $\chi$, at each pixel is calculated via
\begin{equation}
\chi = \frac{1}{2} \text{arctan2}(\frac{U}{Q}),
\end{equation}
where the arctan2 is the four-quadrant arctangent. The positively biased polarization fraction, $P_{\text{frac,b}}$, at each pixel is calculated via
\begin{equation}
P_{\text{frac,b}} = \frac{\sqrt{Q^2+U^2}}{I}.
\end{equation}

Polarization maps have been de-biased in the pipeline via \pfrac\ = $\sqrt{{P_{\text{frac,b}}-\sigma_{P_{\text{frac}}}^2}}$, where $\sigma_{P_{\text{frac}}}$ is the error on $P_{\text{frac,b}}$ (and \pfrac). 

SOFIA HAWC+ is not sensitive to the absolute Stokes parameters, so some amount of spatial filtering via on-the-fly maps affects the data. These can lead to artifacts that show up as unrealistically large \pfrac\ values. However, large \pfrac\ values ($>$20\%) are outside of the areas of interest in this paper, and thus will not be used for any analysis.

Delivered data were in equatorial coordinates, and we rotated them to Galactic coordinates via the python package \texttt{reproject} \citep{Robitaille2020} and properly rotating position angles \citep{Appenzeller1968}. 

\begin{figure*}[ht!]
\begin{center}
\includegraphics[width=2\columnwidth]{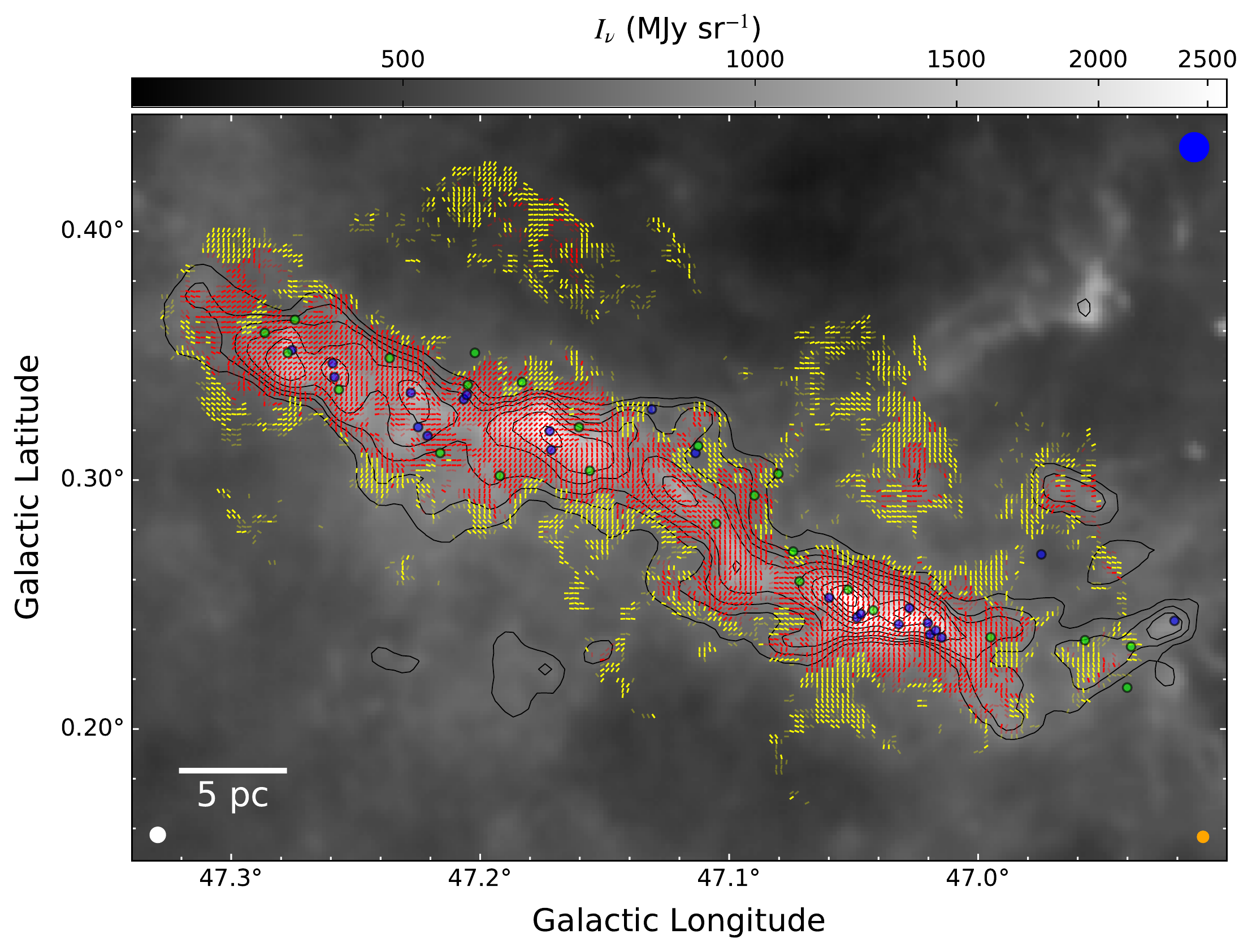}
\end{center}
\caption{Gray-scale $Herschel$ 250\,$\mu$m map of G47 overlaid with inferred B-field vectors from SOFIA HAWC+ Band~E (214\,$\mu$m). Vectors are shown for every two pixels, oversampling the $\sim$5 pixels per beam. Contours show the gas column density (\Ngas), with levels of [0.8, 0.9, 1, 1.2, 1.5, 2, 2.5, 3, 3.5, 4]\,$\times$\,$10^{22}$\,cm$^{-2}$. Red vectors are for polarization fractions (\pfrac) less than 0.15 while yellow vectors are for vectors with \pfrac\ between 0.15 and~1. Light red and yellow vectors indicate \pfrac\ that have signal-to-noise ratios (SNRs) between 2 and 3, while the rest of the vectors have SNRs greater than 3. Blue and green circles indicate locations of Class~I and~II YSOs identified by \citet{ZhangMM2019}. The white, orange, and blue beams (bottom left, bottom right, and top right) show the FWHM resolution for the $Herschel$ 250\,$\mu$m, SOFIA, and \Ngas\ maps respectively.}
\label{fig:G47} 
\end{figure*}

\subsection{Ancillary Data}\label{sec:ancillary}
We use the $Herschel$ 250\,$\mu$m continuum data from Hi-GAL \citep{Molinari2016}, and H$_2$ column density maps (\NHt) generated by \citet{Zucker2018b} from the multi-wavelength Hi-GAL $Herschel$ data. These were subsequently converted to \Ngas\ (i.e., \NHt\,+\,\NHe) by multiplying by the ratio of the mean molecular weight per H$_2$ molecule (\muHt\ = 2.8) divided by the mean molecular weight per particle (\mup\ = 2.37; \citealt{Kauffmann2008}). \citet{Zucker2018b} also fit the ``spine" of the bone -- equivalent to a one-pixel wide representation of its plane-of-the-sky morphology -- using the \texttt{RADFIL} algorithm \citep{Zucker2018c}. The resolution of the column density and spine maps are 43$\arcsec$ ($\sim$1.4\,pc), and they have pixel sizes of 11$\farcs$5\,$\times$\,11$\farcs$5.

We also use \ttcooz\ data from the Galactic Ring Survey \citep[GRS;][]{Jackson2006} and NH$_3$(1,1) data from the Radio Ammonia Mid-plane Survey \citep[RAMPS;][]{Hogge2018}, each of which we convert to velocity dispersion maps via Gaussian fits following \citet{Hogge2018}. While $^{13}$CO(1--0) is detected everywhere along the bone, NH$_3$(1,1) is only detected toward the densest parts. We make a ``final velocity dispersion" map, which we will use to estimate B-field strengths, where we use the NH$_3$(1,1) velocity dispersion when it is available for a particular pixel and \ttcooz\ otherwise. We combine the two together in which we use NH$_3$(1,1) velocity dispersion when it is available for a particular pixel; otherwise we use \ttcooz. In the dense regions where NH$_3$(1,1) is detected, $^{13}$CO(1--0) linewidths tend to be higher (factor of $\sim$2), as it is the combination of diffuse and compact emission. Locations where NH$_3$(1,1) is not detected are expected to be more diffuse, and thus $^{13}$CO(1--0) widths are mostly accurate in these areas.


Locations of the Class~I and~II YSOs were taken from \citet{ZhangMM2019}, which were identified via $Spitzer$ observations. \citet{ZhangMM2019} estimated the survey completeness for Class I YSOs to be a few tenths of a solar mass and for Class II YSOs to be a few solar masses. Class~I YSOs are likely at locations of the highest star formation activity along the bone. \citet{Xu2018} identified several more YSO candidates toward G47, but unlike \citet{ZhangMM2019}, they did not use criteria to exclude contaminants such as AGB stars. 

%
%
%
%
%
%
%
%




\begin{figure*}[ht!]
\begin{center}
\includegraphics[width=2\columnwidth]{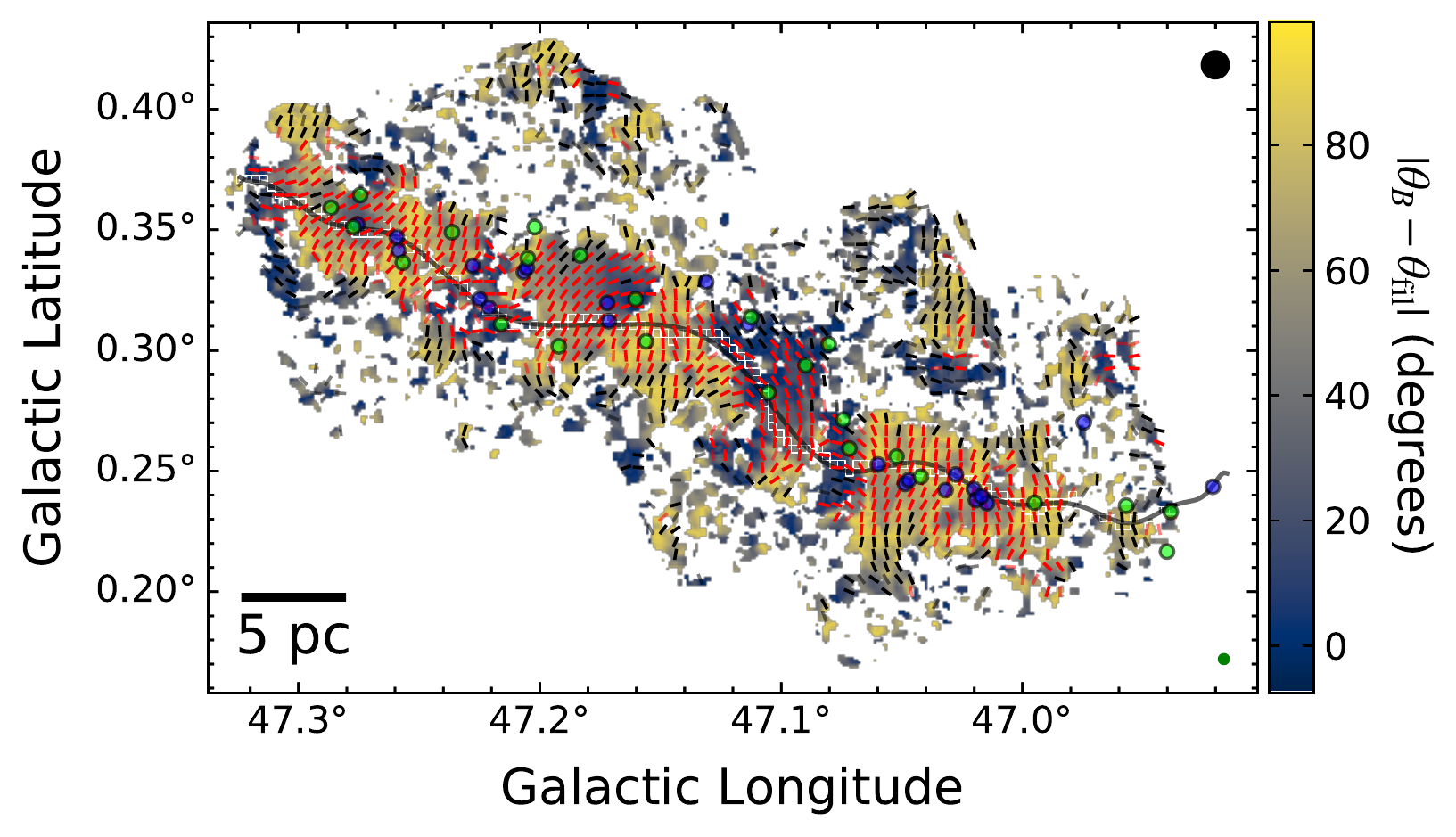}
\includegraphics[width=2\columnwidth]{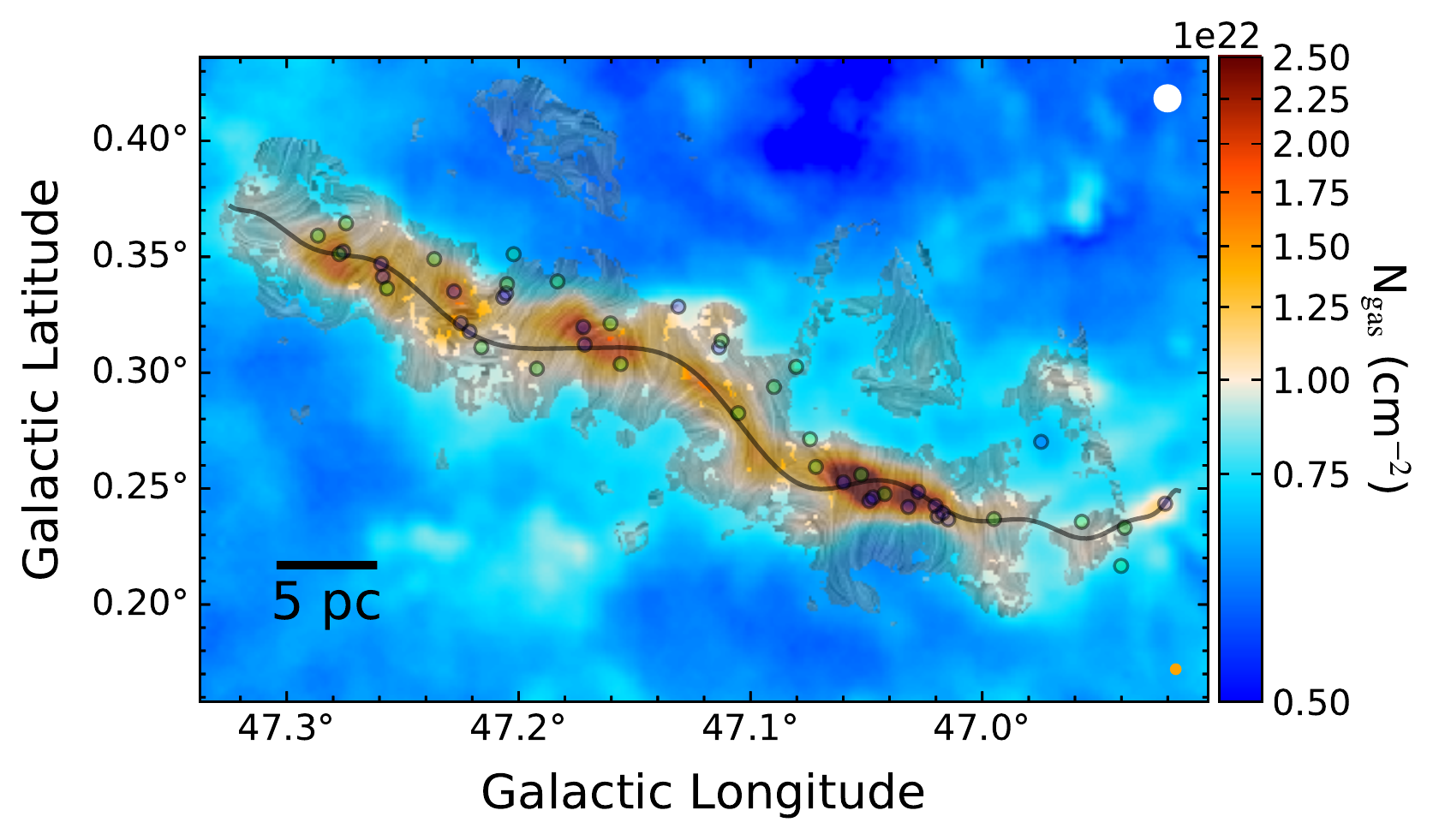}
\end{center}
\caption{{\bf Top Panel:} Magnitude of the difference in angles between the bone direction and B-field direction. White empty squares show the column density spine from \citet{Zucker2018b}, while the black line shows the fitted polynomial to this spine. Red and black vectors are the same as Figure~\ref{fig:G47}'s red and yellow vectors, respectively, except now only one vector is shown for every 5 pixels (or 18$\farcs$5). Blue and green circles indicate locations of Class~I and~II YSOs, respectively, as identified by \citet{ZhangMM2019}. The green beam in the bottom right show the 18$\farcs$2 resolution of SOFIA observations. {\bf Bottom Panel:} LIC map overlaid on a column density map, where the wavy pattern indicates direction of the field. The LIC map filters out pixels for SNR less than 1.5 for \pfrac\ and 10 for Stokes~$I$. For both panels, the beams on the right bottom and top are the FWHM resolution for SOFIA and $Herschel$ spine/\Ngas\ maps, respectively. } 
\label{fig:diff} 
\end{figure*}

\section{Magnetic Field Morphology}\label{sec:results}
Figure~\ref{fig:G47} shows the inferred B-field vectors (i.e., polarization rotated by 90$^\circ$) with \Ngas\ contours overlaid on a $Herschel$ 250~\,$\mu$m map. Immediately evident is the fact that the B-field vectors are not always perpendicular to the filamentary bone. 

To quantify the difference between the position angle (PA; measured counterclockwise from Galactic North) of the B-field and bone's direction, we need to quantify the PA at all locations along the G47's spine. We do this by fitting the spine pixels (Section~\ref{sec:ancillary}) with polynomials of different orders in $l$--$b$ space, and we choose the one with the smallest reduced $\chi^2$. Since the fitted spine pixels are oversampled with 11$\farcs$5 pixels for a 43$\arcsec$ resolution image, we approximate the degrees of freedom to be $\nu$\,=\,(\# of fitted points)/$cf$\,$-$\,(fit order), where correction factor, $cf$, is selected so that the sampling is approximately Nyquist, i.e., $cf =  0.5\times(43\arcsec/11\farcs5) = 1.87$. The best fit polynomial is of 24$^{\text{th}}$ order with a reduced $\chi^2$ of 1.9. For each pixel where we detect polarization, we take the difference between the B-field and the bone PAs by matching them to the closest location to the bone's spine. These differences are shown in top panel of Figure~\ref{fig:diff}. Clearly there are locations along G47 where the field is more perpendicular, more parallel, or somewhere in between. For the two largest column density peaks (see Figure~\ref{fig:G47}), the field is mostly perpendicular. In the bottom panel, we convert vectors to a line integral convolution (LIC) map \citep{Cabral1993} to help visualize the field morphology of G47. The LIC morphology agrees with the analysis presented here.



\begin{figure}[ht!]
\begin{center}
\includegraphics[width=1\columnwidth]{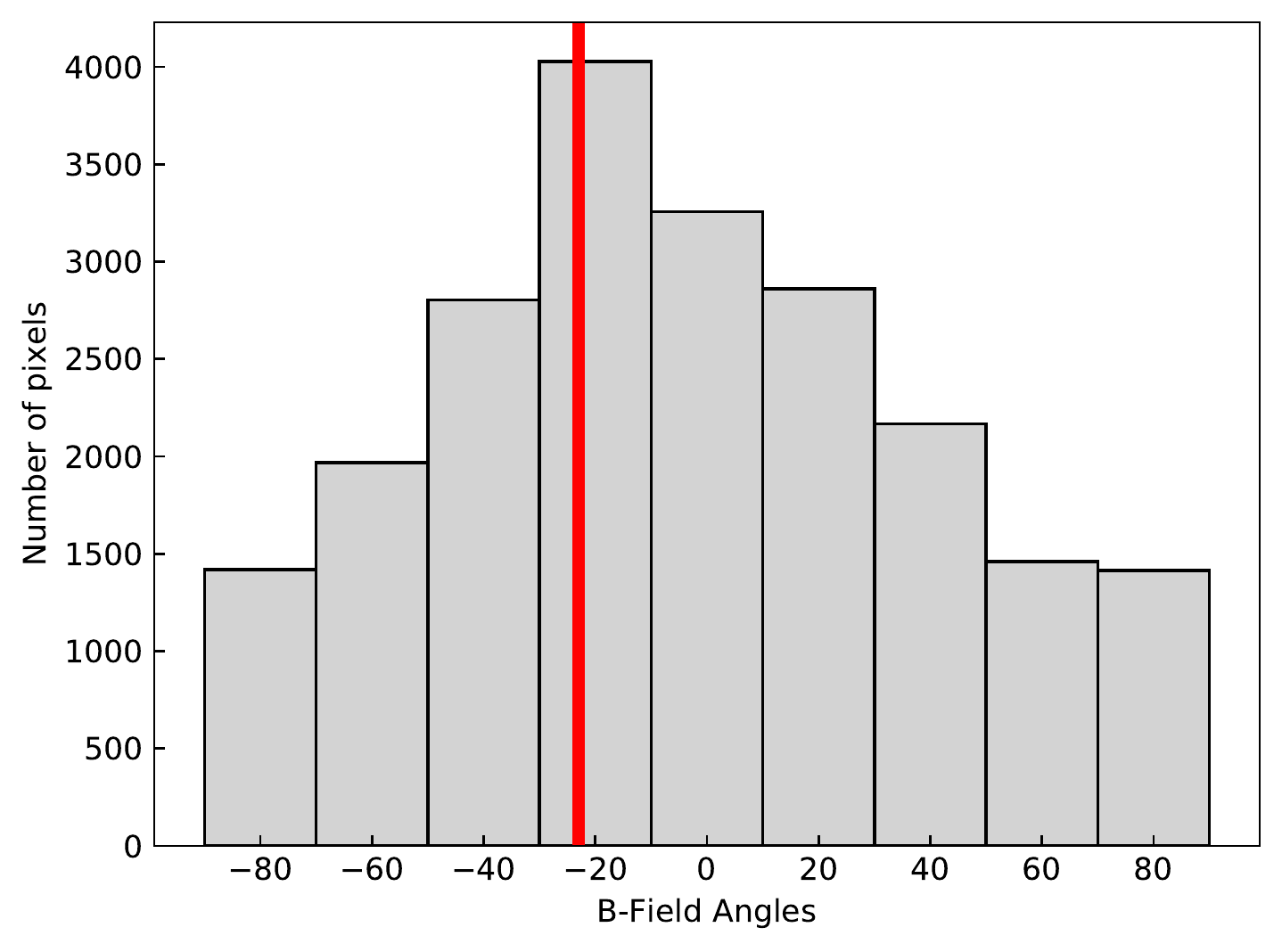}
\end{center}
\caption{Histogram of B-field angle pixels at locations where \Ngas~$>$~$8\times10^{21}$\,cm$^{-2}$. The vertical red line indicates the inferred average B-field angle of $-23^\circ$ across G47, which is based on summing Stokes $Q$ and $U$ with the same column density cutoff.}
\label{fig:Banglehist} 
\end{figure}

We also calculate the average angle across the entire bone by summing Stokes $Q$ and $U$ wherever \Ngas\ is larger than $8\times10^{21}$\,cm$^{-2}$ and converting to a polarization angle. This column density cutoff encompasses the dense elongation of the bone. The polarization PA is 67$^\circ$, or an inferred B-field PA of $-23^\circ$. This angle agrees with the histogram of the B-field angles at locations where \Ngas~$>$~$8\times10^{21}$\,cm$^{-2}$, which is shown in Figure~\ref{fig:Banglehist}. The PA of G47 is about 32$^\circ$ \citep{Zucker2018b}, indicating a difference between the B-field and the angle of the bone of 55$^\circ$. As such, this angle indicates that fields are neither preferentially parallel or perpendicular to the large-scale elongated structure of the bone. {The Galactic field is expected to be along the spiral arms ($b=0^\circ$), and the PA of G47 is also not preferentially parallel or perpendicular to this field.} These findings are consistent with results from \citet{Stephens2011}, which showed that individual star-forming regions are randomly aligned with respect to the Galactic field.


\section{Magnetic Field Estimates}\label{sec:Bfield}
To estimate the plane-of-sky B-field strength ($B_{\text{pos}}$) from polarimetric observations, the Davis-Chandrasekhar Fermi \citep[DCF,][]{Davis1951,Chandrasekhar1953} technique is often used \citep[also see][]{Ostriker2001}. The DCF technique relies on the assumption that turbulent motions of the gas excite Alfvén waves along the magnetic field lines. \citet{Skalidis2021} pointed out that for an interstellar medium that has anisotropic/compressible turbulence, the DCF typically overestimates $B_{\text{pos}}$, and a more accurate expression for the field strength can be derived. This equation, which we will refer to as the DCFST technique, is

\begin{equation}\label{eq:ST} 
B_{\text{pos}} =  \sqrt{2\pi \bar{\rho}}\, \frac{\delta v_{\text{los}}}{\sqrt{\delta \theta}},
\end{equation}
where $\bar{\rho}$ is the average density, $\delta v_{\text{los}}$ is the line of sight velocity dispersion, and $\delta \theta$ is the dispersion in the B-field angles. The classical DCF equation is $B_{\text{pos}} = \mathscr{Q} \sqrt{4\pi \bar{\rho}} \, \delta v_{\text{los}} \,  \delta \theta^{-1}$ where $\mathscr{Q}=0.5$ \citep{Ostriker2001}. As such, $B_{\text{pos,DCFST}}$ is related to the classical $B_{\text{pos,DCF}}$ via $B_{\text{pos,DCFST}} = B_{\text{pos,DCF}}\sqrt{2\delta \theta}$; the expressions for the two $B_{\text{pos}}$ expressions are equal at the Alfv\'enic limit where $\delta \theta = 0.5$. Initial analysis indicates that the DCFST technique more accurately estimates the magnetic field strength than the classical DCF technique \citep{Skalidis2021b}. However, given the potential shortcomings of the DCFST technique \citep{LiPS2021}, it is not yet settled that it is indeed more accurate.

%

We will use this technique in two different ways. First, we calculate the B-field strength across the entire bone by sliding a rectangular box down the spine of the bone and estimating the B-field strength via the DCFST technique for the data in each box. After this, we focus solely on the southwest region, which has the highest column density region and has evidence of a pinched morphology. We fit this morphology using the spheroidal flux freezing (SFF) model outlined in \citet{Myers2018,Myers2020}.

\begin{figure*}[ht!]
\begin{center}
\includegraphics[width=2\columnwidth]{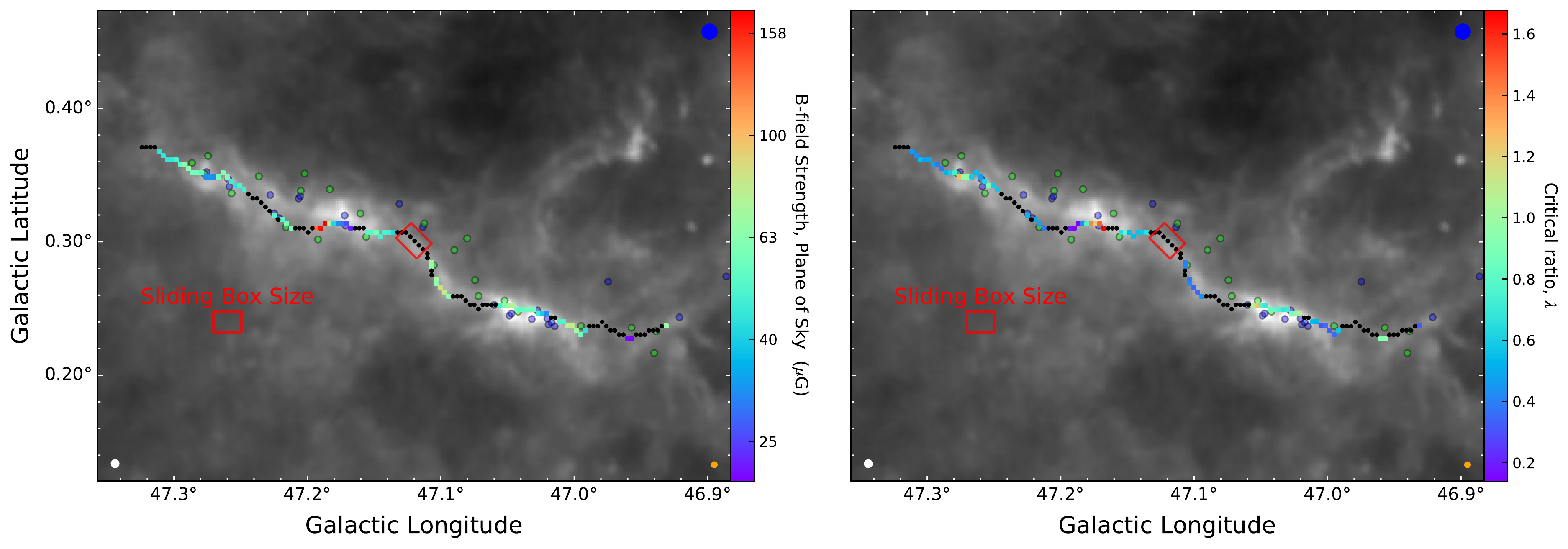}
\end{center}
\caption{Results of the sliding box analysis, where a $74\arcsec \times 55\farcs5$ rectangle was slid across the spine and the B-field was calculated via the DCFST technique. The size of the sliding rectangular box is shown toward the bottom left, with an example of a rectangular box rotated to be along the spine shown at the center of the image. Background for both images are $Herschel$ 250\,$\mu$m with the same stretch (colorbar) and beams as Figure~\ref{fig:G47}.  The left and right panels show the average $B_{\text{pos}}$ and $\lambda$ within the sliding box, respectively. Black circles are locations where we were unable to calculate the field strengths.} 
\label{fig:slidingbox} 
\end{figure*}


\subsection{Sliding box analysis}\label{sec:sliding}
To estimate how the B-field changes across the bone, we apply the DCFST technique along the bone's spine. We do this by ``sliding" a rectangular box down the spine, allowing the box to rotate as the bone's spine change directions in the sky.  The center of the box changes one column density/spine pixel at a time (one spine pixel is 11$\farcs$5\,$\times$\,11$\farcs$5), and the PA of the rectangle is given by the instantaneous slope of the 24$^{\text{th}}$ ordered polynomial fit the spine, as discussed in Section~\ref{sec:results}. The sliding rectangular box has a width, $w$, of 20 HAWC+ pixels (74$\arcsec$) and a height, $h$, of 15 HAWC+ pixels (55$\farcs$5), equivalent to a width and height of $\sim$4 and 3 HAWC+ beams, respectively. These dimensions allow for just over 10 independent beams for each box, which is a sufficient amount of data points for calculating the angular dispersion for the DCFST technique. We do not use a larger box since our underlying assumption is a uniform field in each box, and the field becomes less uniform at larger scales, resulting in an overestimate of the angle dispersion. 

\begin{figure}[ht!]
\begin{center}
\includegraphics[width=1\columnwidth]{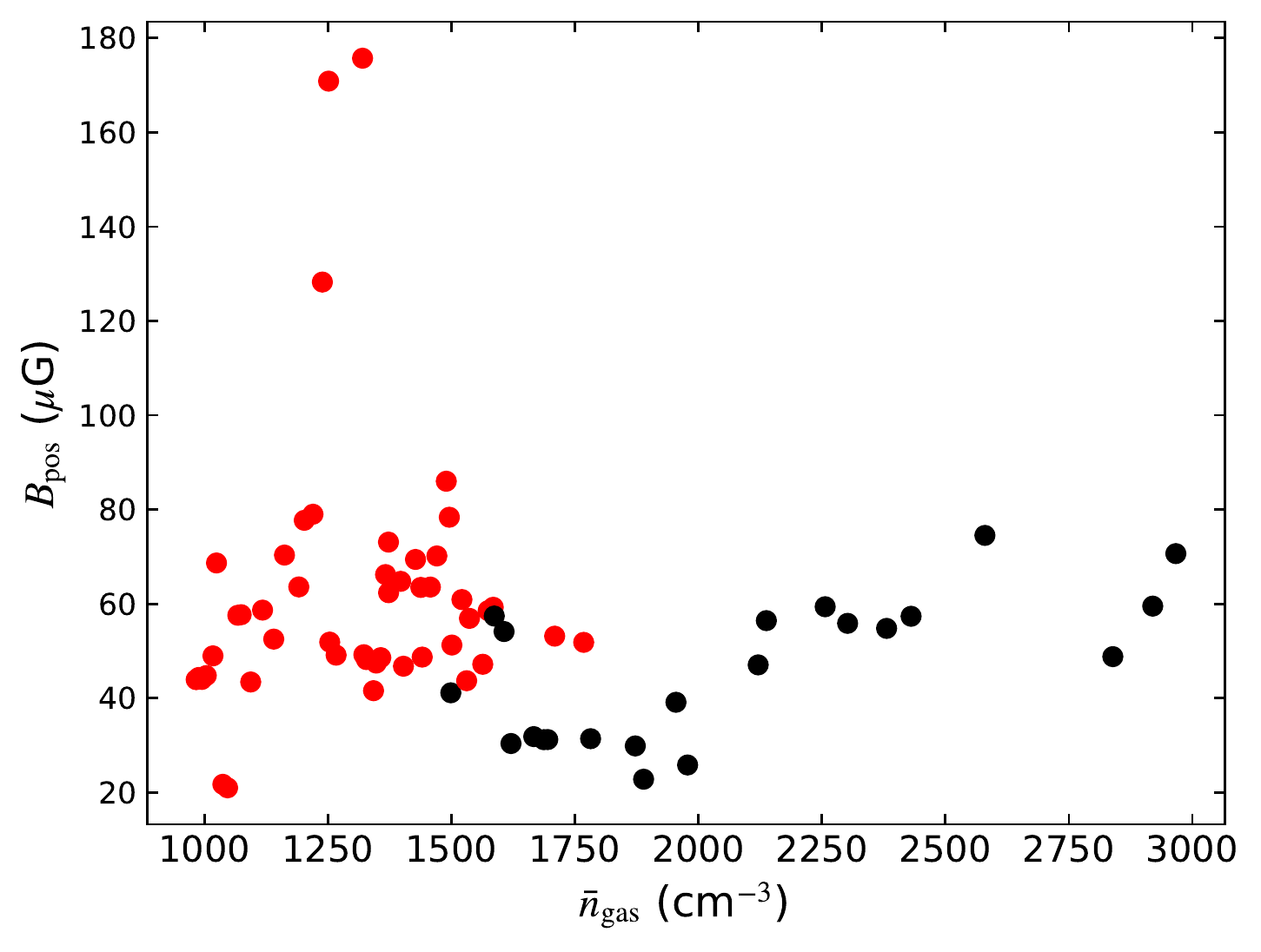}
\includegraphics[width=1\columnwidth]{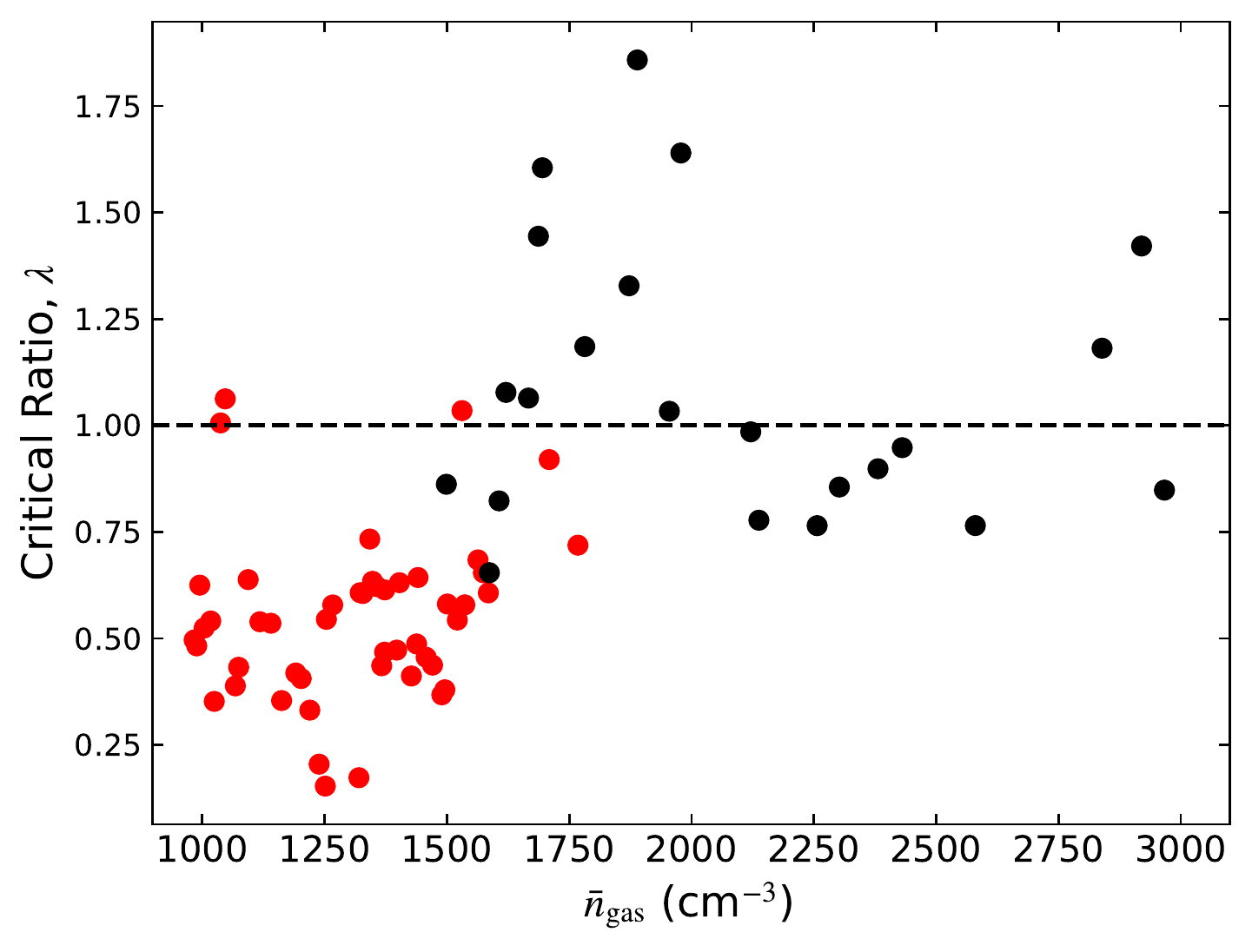}
\end{center}
\caption{Plot of the plane of sky B-field (top panel) and the critical parameter, $\lambda$, (bottom panel) vs the gas column density for each position along the spine for the sliding box analysis. Points are only shown for boxes with $\sigma_\theta < 25^\circ$. Red and black points are for boxes where most the pixels within the box use a velocity dispersion based on $^{13}$CO and NH$_3$(1,1) datasets, respectively. The dashed line is drawn at $\lambda = 1$; values higher than 1 are magnetically supercritical gas, where the B field is weak enough to allow gravitational contraction.}
\label{fig:lambda} 
\end{figure}

We create four image cutouts for each rectangular box sliding down the spine: one for the B-field PA map and another for its error map, one for final velocity dispersion map, and one for the column density map (see Section~\ref{sec:ancillary} for discussion of the latter two). Appendix~\ref{app:box} discusses how to determine whether a given pixel of a map is located within the sliding box. To apply the DCFST technique (Equation~\ref{eq:ST}), we need to estimate $\bar{\rho}$, $\delta v_{\text{los}}$, and $\delta \theta$ for each rectangular box. In the column density cutout, we take the median value as our measure of the mean column density,\footnote{This removes potential outliers. The percent difference between the mean and median for each box is typically less than 5\% and never more than $\sim$10\%} $\bar{N}_{\text{gas}}$, and subsequently convert it to a number density $\bar{n}_{\text{gas}}$ and then $\bar{\rho}$ assuming a cylindrical bone (see Appendix~\ref{app:column}). $\delta v_{\text{los}}$ was chosen to be the median value in the final velocity dispersion cutout. From the B-field PA cutout, we calculate the standard deviation of the cutout, $\delta \theta_{\text{obs}}$, and for its error cutout, we take the median value, which we call $\sigma_{\theta}$. The estimated intrinsic angle dispersion, $\delta \theta$, can be corrected for observational errors such that $\delta \theta = \sqrt{\delta \theta_{\text{obs}}^2-\sigma_{\theta}^2}$. From this we can estimate the plane-of-sky B-field strengths, $B_{\text{pos}}$. We do not calculate the B-fields at locations with $\delta \theta>25^\circ$ \citep{Ostriker2001} since then the turbulence driving the angular dispersion would be super-Alfv\'enic. 

The $B_{\text{pos}}$ map is shown in the left panel of Figure~\ref{fig:slidingbox}, with fields ranging from $\sim$20 to 160\,$\mu$G. We then solve for the critical ratio, $\lambda$, by taking the ratio of the observed and critical mass to magnetic flux ratios, i.e., 
\begin{equation}\label{eq:lambda}
\lambda = \frac{(M/\Phi)_{\text{observed}}}{(M/\Phi)_{\text{crit}}}
\end{equation}
\citep{Crutcher2004}. $\lambda$ parameterizes the relative importance of gravity and magnetic fields.  For $\lambda > 1$, the gas is unstable to gravitational collapse, and when $\lambda < 1$, fields can support the gas against collapse.

The observed mass within the $w\times h$ box is $M_{\text{observed}}$\,=\,\mup$m_p\bar{N}_{\text{gas}} w h$. The magnetic flux is calculated via Appendix~\ref{app:bflux}, and we approximate $(M/\Phi)_{\text{crit}}$ as $1/(2\pi\sqrt{G})$ \citep{McKeeOstriker2007}. The $\lambda$ map is shown in the right panel of Figure~\ref{fig:slidingbox}. Note that errors on these values are difficult to quantify given that some input parameters have non-Gaussian errors, and we make assumptions about the geometry of the bone. These values of $B_{{\text{pos}}}$ and $\lambda$ reflect our best guesses from the data, and we expect them to be correct within a factor of 2--3.  However, since many of the uncertainties in our results are not dominated by random effects but are correlated, for example via the column density or geometrical assumptions, the relative change in these parameters along the bone are likely to be more accurately determined.

Along the spine of the bone, there are two main groups of YSOs: one toward the northeast and one toward the southwest. At both these locations, $\lambda$ tends to be close to or larger than one, indicating that the areas are typically supercritical to collapse. On the other hand, there are several areas along the bone where $\lambda < 1$, indicating that B-fields are potentially strong enough to resist local gravitational contraction. For each position along the spine for which we have a measurement of the B-field, Figure~\ref{fig:lambda} shows $B_{\text{pos}}$ and $\lambda$ as a function of the average number density, $\bar{n}_{\text{gas}}$. The figure also indicates whether most of the velocity dispersion pixels in the sliding box are based on $^{13}$CO or NH$_3$(1,1) since this tracer governs the median velocity dispersion. For densities $\bar{n}_{\text{gas}}$\,$\lesssim$\,1700\,cm$^{-3}$ ($\bar{N}_{\text{gas}}$\,$\lesssim$\,$1.6\times10^{22}$\,cm$^{-2}$), $\lambda$ is typically less than 1 (subcritical), while for higher densities (locations of most YSOs), $\lambda$ is typically higher and often supercritical. Overall, there is little change in $B_{\text{pos}}$ as a function of $\bar{n}_{\text{gas}}$. However, if we only consider the field strengths where NH$_3$ primarily traces the velocity dispersion, i.e., the black points in Figure~\ref{fig:lambda}, the field strength increases slightly as a function of density. Based on the linear regression fit to these points, the slope is 0.022\,$\pm$\,0.006\,$\mu$G/cm$^{-3}$. However, given the dispersion of points over a small range of densities, we cannot draw conclusions from this relation.

Together, these results indicate that the field in some parts of the bone can support against collapse, while in other areas, it is insufficiently strong and thus the gas collapses to form stars. Since YSOs are forming in areas where $\lambda$ is equal to or less than one, this indicates that either we are underestimating $\lambda$ or that the high values of $\lambda$ are more localized to YSOs and would necessitate higher resolution polarimetric observations for proper measurement.


We note that for the $B_{\text{pos}}$ vs  $\bar{n}_{\text{gas}}$ panel, there are 3 points with higher field strengths ($>$100\,$\mu$G) than others. These points are sequentially located next to each other along the spine (see Figure~\ref{fig:slidingbox}). While at these scales G47 has mostly one main velocity component, at this location there appears to be potentially two velocity components that causes the velocity dispersion (and thus the B-field strength) to be overestimated by a factor of $\sim$2. Nevertheless, our fitting routine finds that a one component is slightly better than two, so we only consider it as one component. 


\begin{figure*}[ht!]
\begin{center}
\includegraphics[height=8cm]{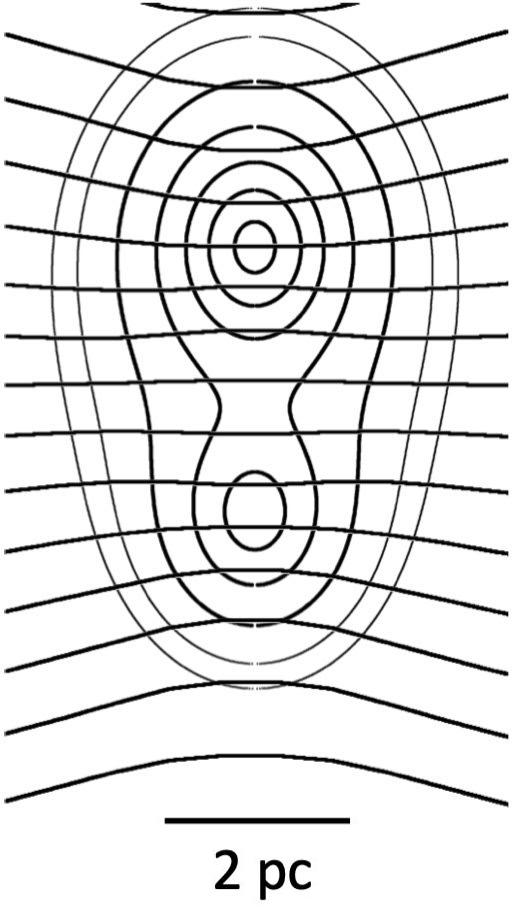}~~~~~~
\includegraphics[height=8cm]{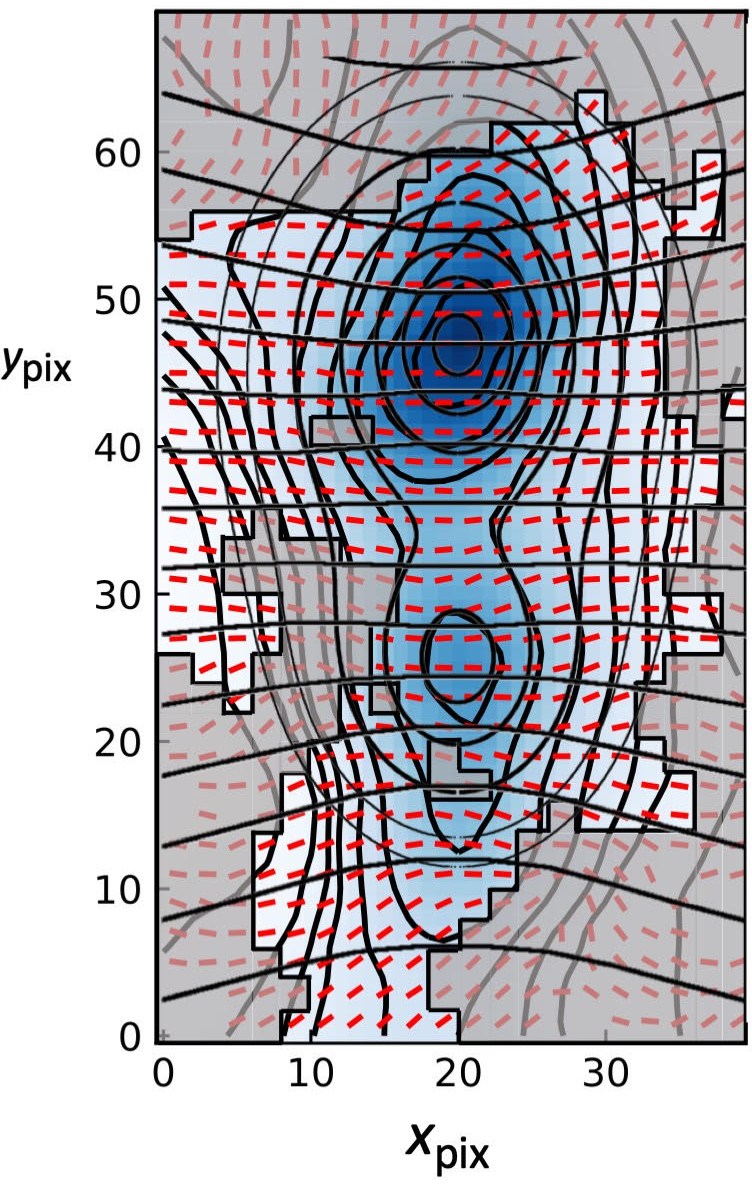}\\
\includegraphics[height=7cm]{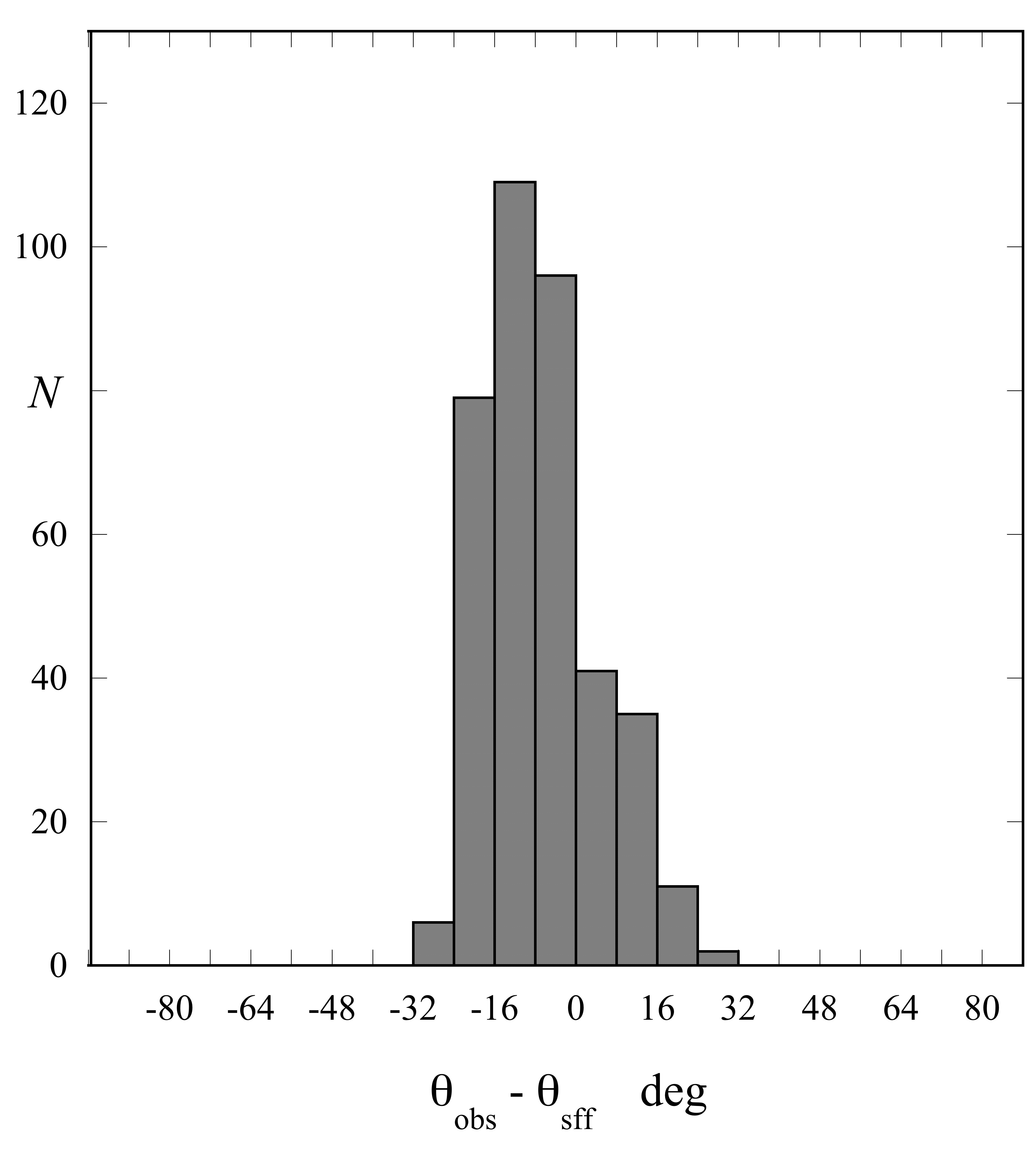}~~~~~~
\includegraphics[height=7cm]{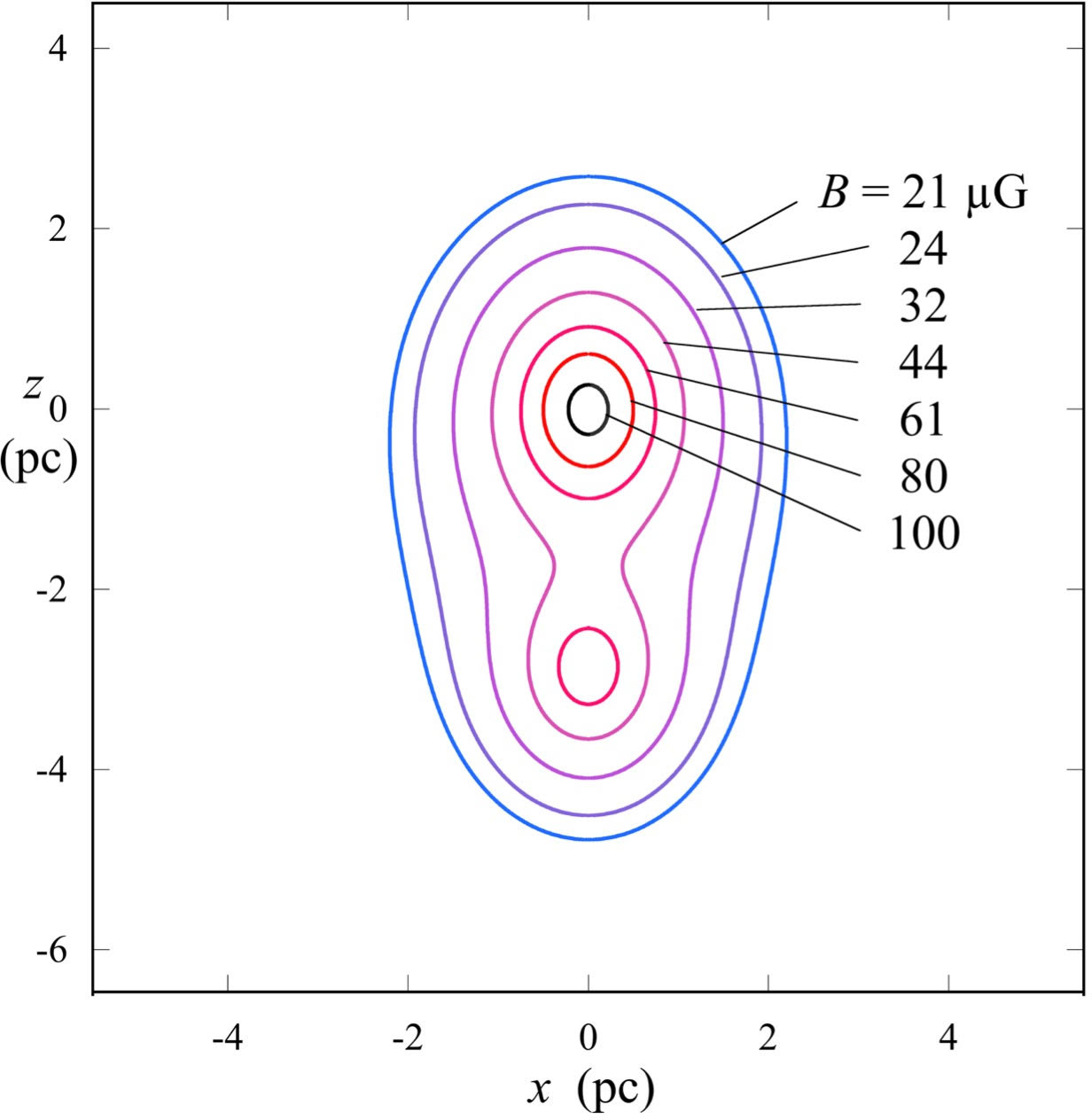}
\end{center}
\caption{Application of the Spheroidal Flux-Freezing (SFF) Technique \citep{Myers2018,Myers2020} to the brightest cores of G47. {\bf Top Left Panel:} Contours of number density $n$ and magnetic field direction in the midplane of the two modeled $p=2$ Plummer spheroids model which have their scale length and aspect ratio fit to the column density map.  The contours are $n/n_0 = 0.08, 0.10, 0.15, 0.25, 0.40, 0.60, 0.90$, assuming negligible background density and peak density $n_0 = 7920$\,cm$^{-3}$. The mostly horizontal magnetic field lines have spacing proportional to $B^{-1/2}$.  {\bf Top Right Panel:} Southwest cores of G47, rotated to be in the coordinate system of the model. Vectors are oversampled and shown for each 3$\farcs$7 pixel. The fit of the top left panel is overlaid on the image. {\bf Bottom Left Panel:} Distribution of difference in angles between observed and modeled magnetic field direction; the standard deviation is $\delta \theta= 11^\circ$. {\bf Bottom Right Panel:} Magnetic field strengths at the same number density contours as the top left panel.
}
\label{fig:SFF} 
\end{figure*}

\subsection{Spheroidal Flux Freezing}
The column density peaks toward the southwest show a pinched B-field morphology. Field lines that are frozen to the gas can create such pinched morphology during collapse. \citet{Mestel1966} and \citet{Mestel1967} calculated the B-field distribution via non-homologous spherical collapse assuming flux-freezing. \citet{Myers2018,Myers2020} extended these calculations for a uniform field collapsing to Plummer spheroids. Since the southwest peaks of G47 have two column density peaks,\footnote{SOFIA 214\,$\mu$m and $Herschel$ 160/250\,$\mu$m maps resolve the top core into two more cores, but for simplicity and to perhaps better reflect the initial collapse, we consider them as one core.} we apply this technique using two Plummer spheroids. We first rotate the delivered G47 data (i.e., in equatorial coordinates) clockwise by 12$^\circ$ to align the column density peaks in the up-down direction. We then fit the column density maps with two $p=2$ Plummer spheroids. With the assumption of an initial uniform field and flux freezing, we can use the resulting Plummer spheroids to predict the field morphology by summing the contributions to horizontal and vertical B-field components from each spheroid (see Sections~2 and~3 of \citealt{Myers2020}). The resulting column density and plane-of-sky B-field lines for the model are shown in the top left panel of Figure~\ref{fig:SFF}. 

We overlay the model on top of the HAWC+ polarimetric map (Figure~\ref{fig:SFF}, top right panel). We mask out angles where the uncertainty in the angle is $>$5$^\circ$. We also mask out angles that differ from the model by more than 25$^\circ$ since these angles are poorly described by the SFF model (i.e., they are outliers that make the distribution non-Gaussian), and these areas may harbor systematic gas flows not included by the model. We calculate the difference in angles between the unmasked inferred field directions and the model (bottom left panel). The dispersion of this distribution is $\delta \theta= 11^\circ$.\footnote{Observational errors, i.e., $\sigma_{\theta}$, in this area are typically only $\sim$2 degrees and thus do not significantly affect $\delta \theta$.}, which is $\sim$75\% of the value of $\delta \theta$ if we did not apply a model. From these data, we can again apply the DCFST technique on the unmasked area. The area is 5.4\,pc$^2$ in size and the median velocity dispersion based on NH$_3$ data is $\delta v_{\text{los}} = 0.8$\,\kms. From our model, we find an average number density of 4300\,cm$^{-3}$.\footnote{The number densities of the SFF analysis are slightly higher than that of the sliding box analysis because the line of sight path length for the SFF model is the width of masked region, which is smaller than the diameter of the bone.}  From these values, we calculate a mean field in the plane of sky in the unmasked area of $B_{\text{pos}}$ = 56\,$\mu$G. The peak total field strength for the SOFIA beam is then $B_0 = 108\,\mu$G, assuming $\bar{B}_{tot}= (4/\pi) B_{{\text{pos}}}$ \citep{Crutcher2004} and $(B_0/\bar{B})=(n_0/\bar{n})^{2/3}$. The total mass in this region is 1170\,$M_\odot$, resulting in a mass-to-flux parameter of $\lambda = 1.7$. The sliding box analysis along the spine of this area found comparable B-field strengths in this region of $\sim$30--75\,$\mu$G with values of $\lambda$ between 0.8 and 1.4. We note the sliding box is $\sim$70\% of the size of the unmasked region. These mass-to-flux ratios lie within the range of mass-to-flux ratios in low-mass star-forming cores, according to a recent study \citep{MyersBasu2021}.

\section{Summary}\label{sec:summary}
We present the first results of the SOFIA Legacy FIELDMAPS survey, which is mapping the B-field morphology across $\sim$10 Milky Way bones.  This initial study focuses on the cloud G47. We find that:
\begin{enumerate}
\item  The plane of sky B-field directions tend to be perpendicular to the projected spine of G47 at the highest mean gas densities of a few thousand cm$^{-3}$, but at lower densities the B-field structure is complex, including parallel and curving directions. 
\item The total inferred B-field across the bone is inconsistent with fields that are parallel or perpendicular to the bone. They are also not aligned with the Galactic plane.
\item We estimate the field strengths using the DCF technique as updated by \citet{Skalidis2021} via two methods: by estimating the B-field within rectangular boxes along G47's spine and by using the SFF technique. We find agreement between the two methods in the area where they both were applied. We find field strengths typically vary from $\sim$20 to $\sim$100\,$\mu$G, but may be up to $\sim$200\,$\mu$G.
\item The spine of G47 has mass to magnetic flux ratios of about 0.2 to 1.7 times the critical value for collapse. Most areas are not critical to collapse. B-fields are thus likely important for support against collapse at these scales in at least some parts of the bones. At the locations of the known YSOs and higher densities, the bone is likely to be more unstable to collapse (i.e., has higher values of $\lambda$). We suspect that high values of $\lambda$ may be more localized with the star formation, necessitating higher resolution polarimetric observations toward the YSOs. 
\end{enumerate}

B-fields likely play a role in supporting the G47 bone from collapse, and they may help shape the bones in areas of highest column density. However, since the field directions for lower column densities are more complex, it is unclear how well B-fields shape or guide flows in the more diffuse areas for the bones. While there are considerable uncertainties in our estimates of the column densities and B-fields, the analysis of the larger sample of bones, available from the full FIELDMAPS survey, will allow more extensive testing of these parameters.

Based on observations made with the NASA/DLR Stratospheric Observatory for Infrared Astronomy (SOFIA). SOFIA is jointly operated by the Universities Space Research Association, Inc. (USRA), under NASA contract NNA17BF53C, and the Deutsches SOFIA Institut (DSI) under DLR contract 50 OK 0901 to the University of Stuttgart. Financial support for this work was provided by NASA through award \#08\_0186 issued by USRA.
CZ acknowledges that support for this work was provided by NASA through the NASA Hubble Fellowship grant \#HST-HF2-51498.001 awarded by the Space Telescope Science Institute, which is operated by the Association of Universities for Research in Astronomy, Inc., for NASA, under contract NAS5-26555.
RJS acknowledges funding from an STFC ERF (grant ST/N00485X/1)
CB gratefully acknowledges support from the National Science Foundation under Award Nos. 1816715 and 2108938.
PS was partially supported by a Grant-in-Aid for Scientific Research (KAKENHI Number 18H01259) of the Japan Society for the Promotion of Science (JSPS). 
ZYL is supported in part by NASA 80NSSC18K1095 and NSF AST-1815784.
LWL acknowledges support from NSF AST-1910364.
We thank Michael Gordon for his effort in setting up the on-the-fly maps for the FIELDMAPS project and Sachin Shenoy for his work on the data reduction.
We thank Miaomiao Zhang for sharing the locations of the YSOs for G47 based on \citet{ZhangMM2019}.
We thank Jin-Long Xu for providing us Purple Mountain Observatory spectral data from \citet{Xu2018}, even though we did not use these data for this Letter.

\facility{SOFIA}
\software{APLpy \citep{Robitaille2012}, 
Astropy \citep{Astropy2013, Astropy2018},
MAGNETAR \citep{Soler2013},
Reproject \citep{Robitaille2020}
}



\appendix
\section{Rectangle Box Analysis}
\subsection{Rectangle for Sliding Box}\label{app:box}
\begin{figure}[ht!]
\begin{center}
\includegraphics[width=1\columnwidth]{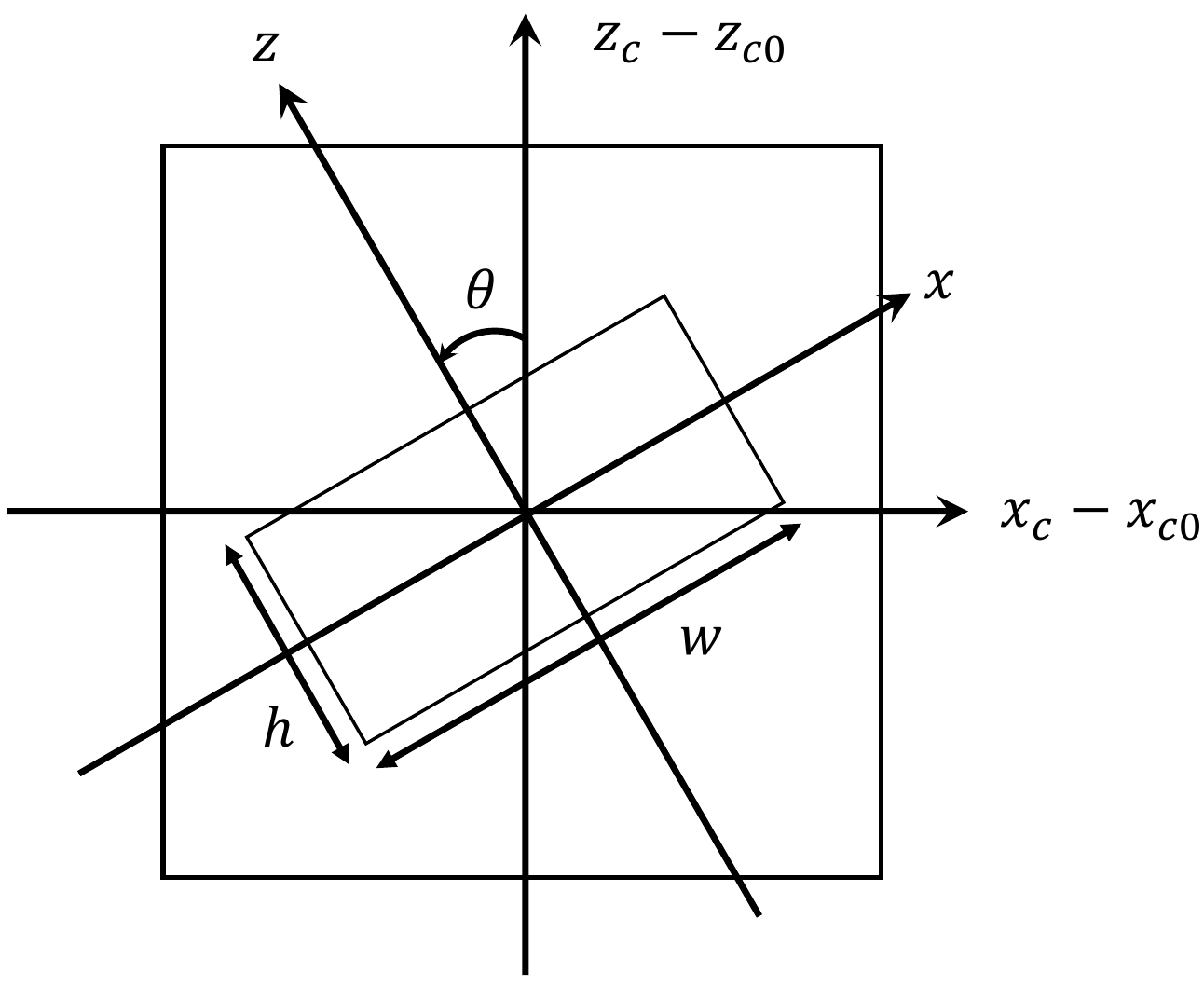}
\end{center}
\caption{Coordinate system used to determine whether pixel $x_c$, $z_c$ is within a rectangle with a PA of $\theta$, a height $h$, and a width $w$ centered at $x_{c0}$, $z_{c0}$. Axes $x$ and $y$ are along the width and height, respectively, while $x_c - x_{c0}$ and $z_c - z_{c0}$ axes are along the left-right and up-down directions, respectively. Diagram is shown for $x_c = x_{c0}$ and $z_c = z_{c0}$. }
\label{fig:rectangle} 
\end{figure}
In Section \ref{sec:sliding}, a rectangular box was moved across the spine of the bone, and we only considered pixels within this box for the DCFST technique. We want to determine whether or not a pixel at location $x_c$, $z_c$ is within a particular rectangle. Consider a rectangle centered at location $x_{c0}$, $z_{c0}$ with height $h$, width $w$, and angle $\theta$ which is measured counterclockwise from up (North), as shown in Figure~\ref{fig:rectangle}. We define one set of axes with respect to the rectangle, where $x$ is in the direction of the width and $z$ is in the direction of the height. We define a second set of axes in the coordinate system of the map (Galactic North-South, West-East for our case), with the axes centered on the pixel of interest relative to the center of the rectangle, i.e., $x_c$\,--\,$x_{c0}$  and $z_c$\,--\,$z_{c0}$. The transformation between the coordinate system is then

\begin{equation}
x = (z_c-z_{c0})\,\text{sin}\,\theta + (x_c - x_{c0})\,\text{cos}\,\theta
\end{equation}
\begin{equation}
z = (z_c-z_{c0})\,\text{cos}\,\theta + (x_c - x_{c0})\,\text{sin}\,\theta
\end{equation}

Pixel $x_c$, $z_c$ will be inside the rectangle if $|x| \leq\ w/2$ and $|z| \leq\ h/2$. All pixels meeting these criteria in a given SOFIA  map are considered within a particular rectangular box. For our analysis, we used $w = 20\,\text{pixels} = 74\arcsec$ and $h = 15\,\text{pixels} = 55\farcs5$. In our particular case, the chosen pixel size oversamples  the beam. However, oversampling does not change the true dispersion, mean, or median, and potentially can estimate these parameters more accurately since at least Nyquist sampling is needed to capture all features of a map.


\subsection{Inferring \smallngas\ and $\rho$ from \Ngas}\label{app:column}
%
We want to calculate the volume density for the sliding box to apply the DCFST method. Consider a cylinder (approximation for a bone or filament) with the x-axis along the length of the cylinder, the z-axis perpendicular to the x-axis and in the plane of sky, the y-axis along the line of sight, and a center at $(x,y,z) = (0,0,0)$. This box has a height $h$ (Figure~\ref{fig:rectangle}), which we define to extend from $-z_1$ to $z_1$ so that $z_1$ = $h/2$. If the observed mean column density within the box, $\bar{N}(z_1,R)$, is calculated, then the mean number density within the box, $\bar{n}(z_1,R)$, is

\begin{equation}
\bar{n}(z_1,R) = \frac{\bar{N}(z_1,R)}{2\bar{Y}(z_1,R)}
\end{equation}
where $2\bar{Y}(z_1,R) = (1/z_1)\int_{0}^{z_1} dz\sqrt{R^2-z^2}$ is the path length averaged over the heights 0 to $z_1$. Substituting the path length in the above equation, we arrive at the final equation for $\bar{n}(z_1,R)$ of

\begin{equation}
\bar{n}(z_1,R) = \frac{\bar{N}(z_1,R)}{\sqrt{R^2-z_1^2}+\frac{R^2}{z_1} \rm{tan}^{-1}\frac{z_1}{\sqrt{R^2-z_1^2}}}.
\end{equation}
 For G47, we take $R=1.6\,\text{pc}$ \citep{Zucker2018b} and $h = 2z_1 = 15\,\text{pixels} = 55\farcs5$. To convert to the mean mass volume density, $\bar{\rho}$, $\bar{n}$ should be multiplied by the mean molecular weight times the hydrogen mass, $m_H$. If $\bar{n}$ is the gas volume density (used in this study), the mean molecular weight per particle, $\mu_p = 2.37$, should be used; if $\bar{n}$ is the H$_2$ volume density, then the mean molecular weight per H$_2$ molecule, \muHt\,=\,2.8, should be used \citep{Kauffmann2008}.

\begin{figure*}[ht!]
\begin{center}
\includegraphics[width=2\columnwidth]{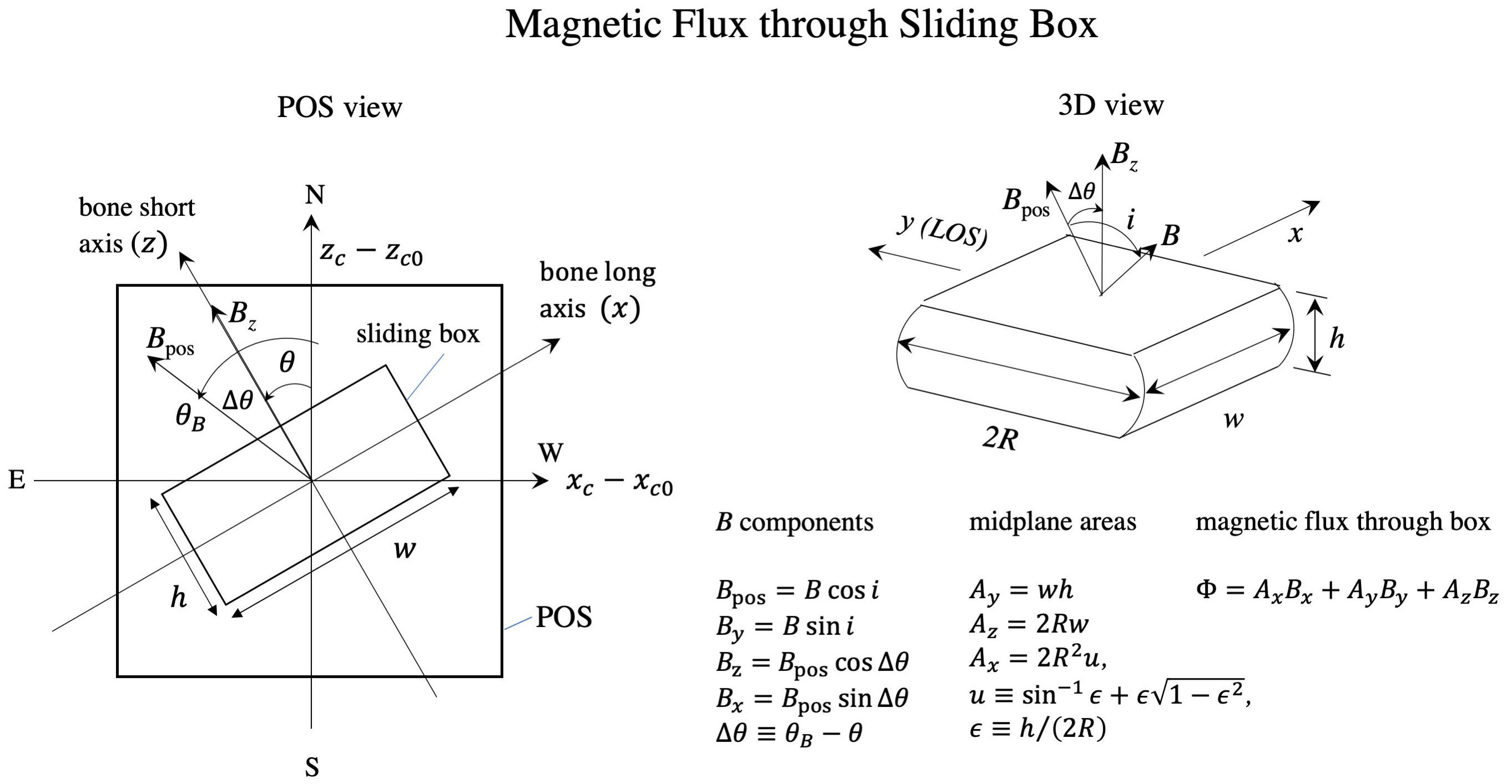}\label{fig:magneticflux}
\end{center}
\caption{Graphic illustrating how to calculate the magnetic flux for the sliding rectangular box technique. }
\end{figure*}
\subsection{Magnetic Flux}\label{app:bflux}
We draw the sliding rectangular box we use in Figure~\ref{fig:magneticflux} from the plane-of-sky perspective (left panel) and a 3-dimensional view (right panel). The x, y, and z axes are along the bone's short axis, the line of sight (LOS), and the bone's long axis, respectively. The magnetic flux is simply the sum of the flux in each dimension, i.e.,

\begin{equation}
\Phi = A_x B_x + A_y B_y + A_z B_z
\end{equation}
where $A_i$ and $B_i$ are the areas and B-fields for each axis direction. We measure the $B_{\text{pos}}$ from the DCFST method, and we have a typical B-field direction, $\theta_B$, which we take to be the median angle in the sliding box analysis. The true B-field direction has an inclination $i$ along the line of sight ($B_{\text{pos}} = B$ for $i = 0$). For a rectangular box rotated so that the z-axis has a PA of $\theta$, we define $\Delta \theta \equiv \theta_B - \theta$. The solutions for the parametrization for each $A_i$ and $B_i$ are indicated in Figure~\ref{fig:magneticflux}.  The formula for $A_x$ accounts for the fact that it is the crosssectional area of a circular cylinder with parallel planar sides. The inclination $i$ is unknown. The true B-field is typically chosen based on a statistical average with inclination such that $B_{\text{pos}}/B = \pi/4$ \citep{Crutcher2004}. We choose $i$ to be based on this average so that $i = i_{\text{avg}}= \text{cos}^{-1}(\pi/4) \approx 38.2^\circ$. For a given bone radius $R$ and $w \times h$ rectangular box, we can now calculate $\Phi$ from the measurements of $B_{\text{pos}}$ and $\theta_B$ in each box. 

For a given field strength and orientation, the magnetic flux depends on the geometry of the smoothing box since the projected box area in the field direction varies with the relative box dimensions and with the box inclination. As such, the measurements should be placed in the context of the box dimensions and field direction. Nevertheless, for the adopted box dimensions, changing box orientation by 90$^\circ$ changes the derived flux by less than 25\%. For a typical box orientation, reducing the adopted box width $w$ by a factor of 2 reduces the derived flux by $\sim$35\%.

\clearpage
\bibliography{stephens_bib}

\end{document}